\documentclass[12pt]{article}
\usepackage{amsfonts}
\usepackage{amsmath}
\usepackage{amssymb}
\usepackage{graphicx}
\usepackage{color}
\usepackage[all, knot]{xy}
\usepackage{tikz}

\usepackage{ulem}

\usepackage{hyperref}
\usepackage[utf8]{inputenc}
\usepackage{epstopdf}
\usepackage[footnotesize]{caption}
\usepackage{amsthm}
\usepackage{enumitem}
\usepackage{mathrsfs}

\usepackage[margin=3cm]{geometry}
\def \be {\begin{equation}}
\def \ee {\end{equation}}
\def \bea {\begin{eqnarray}}
\def \eea {\end{eqnarray}}
\def \nn {\nonumber}

\def \rr {\raise.35ex\hbox{\small $\prime$}\kern-.17em{\mbox{\large $\imath$}}}

\def \dels {\partial\kern-.6em /\kern.1em}
\def \As {{A\kern-.5em / \kern.5em}}
\def \Ds {D\kern-.7em / \kern.5em}

\def \ks {k\kern-.5em /}
\def \ls {l\kern-.5em /}

\newcommand{\ci}[1]{}

\newcommand{\ba}{\begin{eqnarray}}
\newcommand{\ea}{\end{eqnarray}}
\newcommand{\bal}{\begin{align}}
\newcommand{\eal}{\end{align}}
\newcommand{\bay}[1]{\left(\begin{array}{#1}}
\newcommand{\eay}{\end{array}\right)}

\newcommand{\hide}[1]{}

\newlist{axioms}{enumerate}{2}
\setlist[axioms,1]{label=\textbf{A\arabic{axiomsi}.}, ref=A\arabic{axiomsi}}
\setlist[axioms,2]{label=\textbf{A\arabic{axiomsi}\rlap{\myEnumCounter{axiomsii}}.},%
                   ref=A\arabic{axiomsi}\myEnumCounter{axiomsii},%
                   align=parleft,%
                   leftmargin=0em,%
                   itemsep=1.4ex,%
                   before={\stepcounter{axiomsi}}}

  \usetikzlibrary{decorations.markings}

\begin{document}

\begin{titlepage}
\begin{center}

\textbf{\LARGE
Study of Asymptotic Free Scalar Field Theories from\\ 
Adaptive Perturbation Method 
\vskip.3cm
}
\vskip .5in
{\large
Chen-Te Ma$^{a,b,c,d,e,f}$ \footnote{e-mail address: yefgst@gmail.com} 
and Hui Zhang$^{c,e,g,h}$ \footnote{e-mail address: mr.zhanghui@m.scnu.edu.cn}
\\
\vskip 1mm
}
{\sl
$^a$
Department of Physics and Astronomy, 
Iowa State University, Ames, Iowa 50010, US.
\\
$^b$
Asia Pacific Center for Theoretical Physics,\\
Pohang University of Science and Technology, 
Pohang 37673, Gyeongsangbuk-do, South Korea. 
\\
$^c$
Guangdong Provincial Key Laboratory of Nuclear Science,\\
 Institute of Quantum Matter,
South China Normal University, Guangzhou 510006, Guangdong, China. 
\\ 
$^d$ 
School of Physics and Telecommunication Engineering,\\
South China Normal University, Guangzhou 510006, Guangdong, China. 
\\ 
$^e$
Guangdong-Hong Kong Joint Laboratory of Quantum Matter,\\
 Southern Nuclear Science Computing Center, 
South China Normal University, 
Guangzhou 510006, Guangdong, China. 
\\
$^f$ 
The Laboratory for Quantum Gravity and Strings,\\
 Department of Mathematics and Applied Mathematics, 
University of Cape Town,
 Private Bag, Rondebosch 7700, South Africa.
 \\
 $^g$
 Key Laboratory of Atomic and Subatomic Structure and Quantum Control,\\ 
Institute of Quantum Matter, South China Normal University, Guangzhou 510006, Guangdong, China.
\\
$^h$
Physics Department and Center for Exploration of Energy and Matter, Indiana University, Bloomington, Indiana 47408, US.
}\\
\vskip 1mm
\vspace{40pt}
\end{center}

\newpage
\begin{abstract} 
We focus on the behavior of (2+1)d $\lambda\phi^4$ and (5+1)d $\lambda\phi^3$ or $\lambda|\phi|^3$ theories in different regimes and compare the results obtained from the adaptive perturbation method with those obtained from lattice simulation.
These theories are simple models that exhibit asymptotic freedom, which is a property that is also observed in more complex theories such as QCD, which describes the strong interaction between quarks and gluons.
Asymptotic freedom is an important feature of these theories because it allows for a perturbative treatment of interactions at high energies.
However, the standard perturbation scheme breaks down in the presence of strong interactions, and the adaptive perturbation method, which involves resuming the Feynman diagrams, is more suitable for studying these interactions.
Our research involves comparing the perturbation result to lattice simulation.
In the case of the $\phi^3$ theory, there is no stable vacuum, so we explore evidence from the $|\phi|^3$ theory instead.
Our results appear to show that resummation improves the strong coupling result for both the $\lambda\phi^4$ and $\lambda|\phi|^3$ theories.
Additionally, we improve the resummation method for the three-point coupling vertex and study the RG flow to analyze the resummation contribution and theoretical properties.
\end{abstract}
\end{titlepage}

\section{Introduction}
\label{sec:1}
\noindent
Asymptotic freedom  \cite{Gell-Mann:1954yli,Gross:1973id,Politzer:1973fx,Luscher:1987ay,Aizenman:1981du} is a property of Quantum Field Theory (QFT) \cite{Goldstone:1961eq,Feynman:1963uxa,Ma:2018efs}, in which the strength of the interaction between particles decreases as the energy scale at which they are probed increases.
Quantum Chromodynamics (QCD) is a theory that exhibits asymptotic freedom.
At low energies, the interactions become strong, and non-perturbative techniques are needed to understand the behavior of theory.
\\

\noindent
The standard perturbation method is indeed limited in its ability to explore strongly coupled systems, as it relies on a small coupling constant to expand the perturbation series \cite{Hioe:1978jj}.
The unperturbed part of the system corresponds to the non-interacting sector, where the Hamiltonian can be easily diagonalized.
This approach is useful for weakly coupled systems, but it becomes problematic in the strongly coupled regime.
To address this issue, the adaptive perturbation method was introduced \cite{Weinstein:2005kw,Weinstein:2005kx}.
This approach involves introducing an adaptive parameter into the Hamiltonian and optimizing it to minimize the energy of the system \cite{Weinstein:2005kw,Weinstein:2005kx}.
The advantage of this method is that it can be applied to a wide range of systems, including strongly coupled ones.
In the adaptive perturbation method, the unperturbed part of the system is still solvable because of choosing an equal number of the annihilation ($a$) and creation operators ($a^{\dagger}$) \cite{Bargmann:1962zz}.
The choice of the parameter is determined by minimizing the energy of the system, which leads to an improved perturbation result.
\\

\noindent
While the adaptive perturbation method has been successfully applied to Quantum Mechanics \cite{Ma:2019pxd,Ma:2020ipi,Ma:2020syr}, its application to QFT requires more care, particularly for Lorentz symmetry.
Nonetheless, recent studies have shown promising results in extending the adaptive perturbation method to QFT, such as the (3+1)d $\lambda\phi^4$ theory \cite{Ma:2022atx}.
The adaptive perturbation method is a useful extension of the standard perturbation method that allows for the study of strongly coupled systems.
\\

\noindent
The expansion of the scalar field is \cite{Ma:2022atx}
\bea
\phi(x)|0\rangle&\equiv& \bigg(\int \frac{d^3\vec{k}}{(2\pi)^3}\frac{1}{\sqrt{2\gamma(\vec{k})}}(a_{\vec{k}}e^{ikx}+a^{\dagger}_{\vec{k}}e^{-ikx})
\bigg)\bigg|0\bigg\rangle,
\nn\\
\Pi(x)|0\rangle&\equiv&\partial_0\phi(x)|0\rangle
\label{sm}
\eea
with the adaptive parameter $\gamma^2(\vec{k})=\vec{k}^2+\gamma^2(0)$ and $kx\equiv-\gamma(\vec{k})x^0+\vec{k}\cdot\vec{x}$.
The $\Pi$ is the canonical momentum.
The extreme value of the ground state energy determines $\gamma^2(0)$.
In general, the expansion from the plane wave is not proper for the interacting part.
Here we change the basis by the Bogoliubov transformation: $U(\theta)a_{\vec{k}}U^{-1}(\theta)\rightarrow a_{\vec{k}}$ and $U(\theta)|0\rangle\rightarrow|0\rangle$ to implement the plane wave expansion.
The expectation value is invariant under the transformation.
Because the commutation relation remains the same, it allows the appearance of the adaptive parameter in the expansion.
Indeed, the expansion does not apply to the field but is applicable when acting on a state \cite{Ma:2022atx}.
We can examine the expansion by choosing the un-perturbed Hamiltonian $H_0$ with an equal number of $a_{\vec{k}}$ and $a^{\dagger}_{\vec{k}}$ \cite{Ma:2022atx}.
Then the time evolution of fields is the same as in quantizing a non-interacting theory \cite{Ma:2022atx}.
Hence the plane wave expansion is proper for the combination of field and states \cite{Ma:2022atx}.
The result also implies that the state is the ground state of $H_0$ but not the non-interacting sector \cite{Ma:2022atx}.
Now the particle spectrum of $H_0$ is $\sqrt{\vec{k}^2+\gamma^2(0)}$.
Therefore, it is convenient to consider the un-perturbed Lagrangian \cite{Ma:2022atx}
\bea
{\cal L}_0=-\frac{1}{2}(\partial_{\mu}\phi\partial^{\mu}\phi+\gamma^2(0)\phi^2).
\label{unperturbed}
\eea
The Lorentz symmetry is manifest \cite{Ma:2022atx}, and the leading-order result remains the same.
The $\gamma^2(0)$ leads to a resummation of bubble diagrams \cite{Ma:2022atx}.
The behavior of Asymptotic Scalar Field Theories from the adaptive perturbation method can be studied using the resummation technique.
The adaptive perturbation method involves the introduction of an adaptive parameter in the expansion of the scalar field, which is determined by the extreme value of the ground state energy.
This allows for the resummation of bubble diagrams and the use of Feynman diagram techniques to study the perturbation \cite{Ma:2022atx}.
If we only consider the resummation of the two-point function up to the one-loop diagram, it is equivalent to HTL perturbation theory \cite{Andersen:2002ey}.
However, the adaptive perturbation method can have the resummation of loop diagrams for the one-point and higher-point functions in general, and it will show the difference from the HTL perturbation theory.
\\

\noindent
In the strongly coupled regime, the computation of correlation functions \cite{Schwinger:1951ex} using this technique works well, as it has been shown by comparing it to lattice simulations \cite{Ma:2022atx}.
This suggests that the resummation technique has a general application and can be used to study the behavior of asymptotic scalar field theories.
The central question that we would like to address in this letter is the following: {\it What is the behavior of Asymptotic Scalar Field Theories from Adaptive Perturbation Method?}
\\

\noindent
We study the properties of asymptotic free scalar fields \cite{Symanzik:1973hx,Kleefeld:2005hf,Romatschke:2023sce}, using the (2+1)d $\lambda\phi^4$ and (5+1)d $\lambda\phi^3$ or $\lambda|\phi|^3$ theories as examples.
The fact that the $\lambda\phi^3$ does not have a stable vacuum makes it difficult to study in lattice simulations, so we choose to study the $|\phi|^3$ term instead.
We then apply the adaptive perturbation method to both the $\phi^4$ and $|\phi|^3$ theories and find that the results of the correlation functions match well with the lattice simulations, even for the strongly coupled regime.
Next, we improve the resummation method for the three-point coupling vertex.
Finally, we study the Renormalization Group (RG) flow to investigate the properties of the asymptotic free scalar field.

\section{Adaptive Perturbation Method}
\label{sec:2} 
\noindent 
We apply the adaptive perturbation method \cite{Ma:2022atx} to (2+1)d $\lambda\phi^4$ and (5+1)d $\lambda|\phi|^3$ theory. 
For $\phi^4$ theory, we calculate $\langle \phi\rangle$, $\langle\phi^2\rangle$, and $\langle\phi^4\rangle$ for $\lambda=1, 2, 4$. 
We calculate $\langle\phi\rangle$ and $\langle\phi^2\rangle$ for $\lambda=0.25, 0.5, 1, 2$ in $|\phi|^3$ theory. 
For each case, we obtain a nice comparison to the lattice simulation. 

\subsection{$\phi^4$ Theory}
\noindent 
The adaptive perturbation method sums the bubble diagrams for the four-scalar interaction theory \cite{Ma:2022atx}. 
This resummation introduces the modified propagator and also the condensation $\phi_0$ \cite{Ma:2022atx}. 
The result applies well to the (3+1)d case \cite{Ma:2022atx}. 
However, a diagrammatic method is independent of the spacetime dimensions. 
Hence we can also apply this method to (2+1)d $\lambda\phi^4$ theory \cite{Ma:2022atx}.  
\\

\noindent 
The Euclidean Lagrangian for $\phi^4$ theory is 
\bea
{\cal L}_4=\frac{1}{2}\partial_{\mu}\phi\partial^{\mu}\phi
+\frac{m^2}{2}\phi^2
+\frac{\lambda}{4!}\phi^4,  
\eea
where $m$ and $\lambda$ are the bare mass and coupling constant, respectively. 
Now we introduce the constant condensation $\phi_0=\phi-\phi_f$, and it introduces the momentum dependence to the renormalized mass $\gamma$. 
The square of the renormalized mass $\gamma^2(k_E)$ and the constant condensation $\phi_0$ are given by: 
\bea
&&
\gamma^2(k_E)
\nn\\
&=&m^2+\frac{\lambda}{2}\phi_0^2
+\frac{\lambda}{2}\int\frac{d^3q_{E}}{(2\pi)^3}\ \frac{1}{q_{E}^2+\gamma^2(q_E)}
\nn\\
&&
-\frac{\lambda^2\phi_0^2}{2}
\int\frac{d^3q_E}{(2\pi)^3}\ \frac{1}{q_E^2+\gamma^2(q_E)}
\nn\\
&&\times
\frac{1}{(k_E-q_E)^2+\gamma^2(k_E-q_E)};
\eea
\bea
\phi_0\bigg(
 m^2+\frac{\lambda}{3!}\phi_0^2+\frac{\lambda}{2}\int\frac{d^3k_E}{(2\pi)^3}\ \frac{1}{k_E^2+\gamma^2(k_E)}\bigg)=0, 
 \eea 
where $k_E$ is the Euclidean momentum. 
It is more convenient to show the unperturbed Lagrangian on the momentum space 
\bea
{\cal L}_{k, 0}=\frac{1}{2}\int\frac{d^3k_E}{(2\pi)^3}\big(k_E^2+\gamma^2(k_E)\big)\tilde{\phi}_f(k_E)\tilde{\phi}_f(-k_E),  
\eea
where $\tilde{\phi}_f$ is the fluctuation field on the momentum space 
\bea
\phi_f(x_E)=\int\frac{d^3k_E}{(2\pi)^3}\ e^{ik_Ex_E}\tilde{\phi}_f(k_E). 
\eea
\\

\begin{figure}
\begin{center}
    \includegraphics[width=0.49\textwidth]{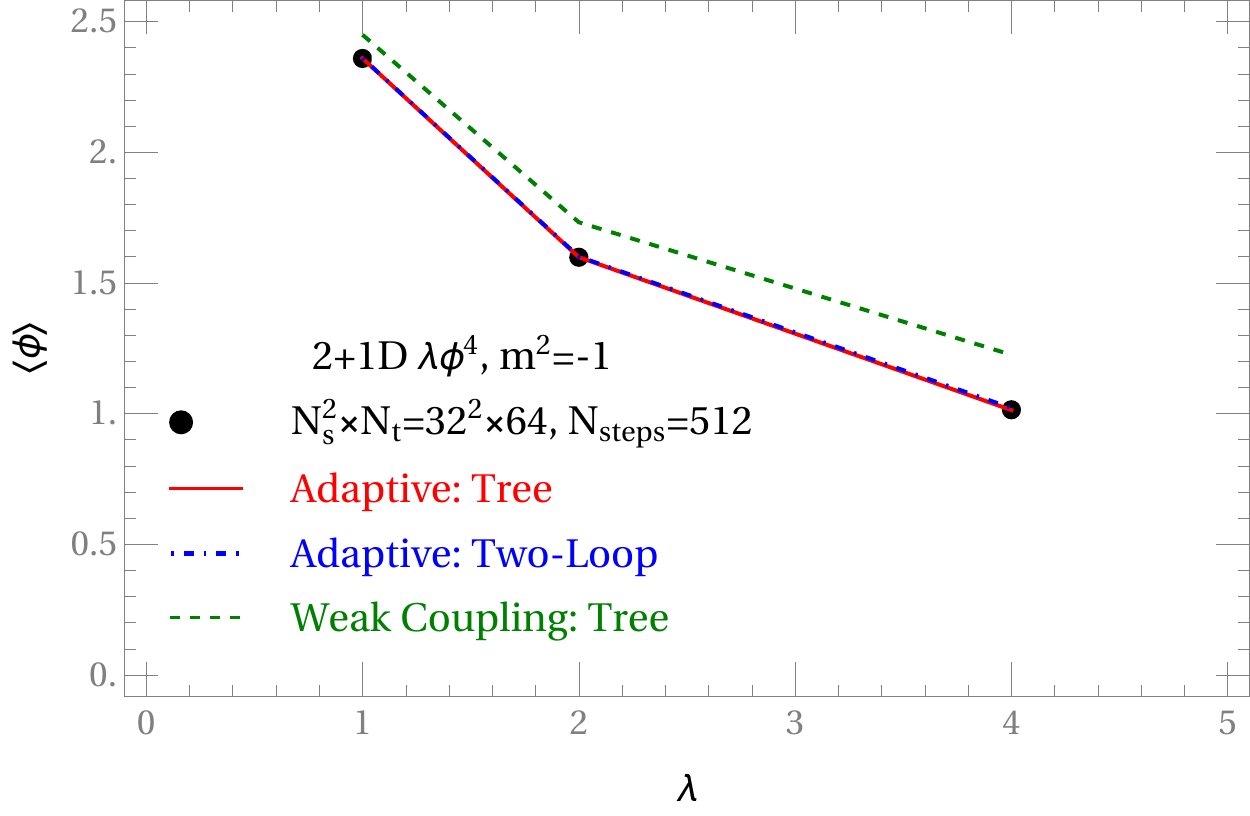}
    \includegraphics[width=0.49\textwidth]{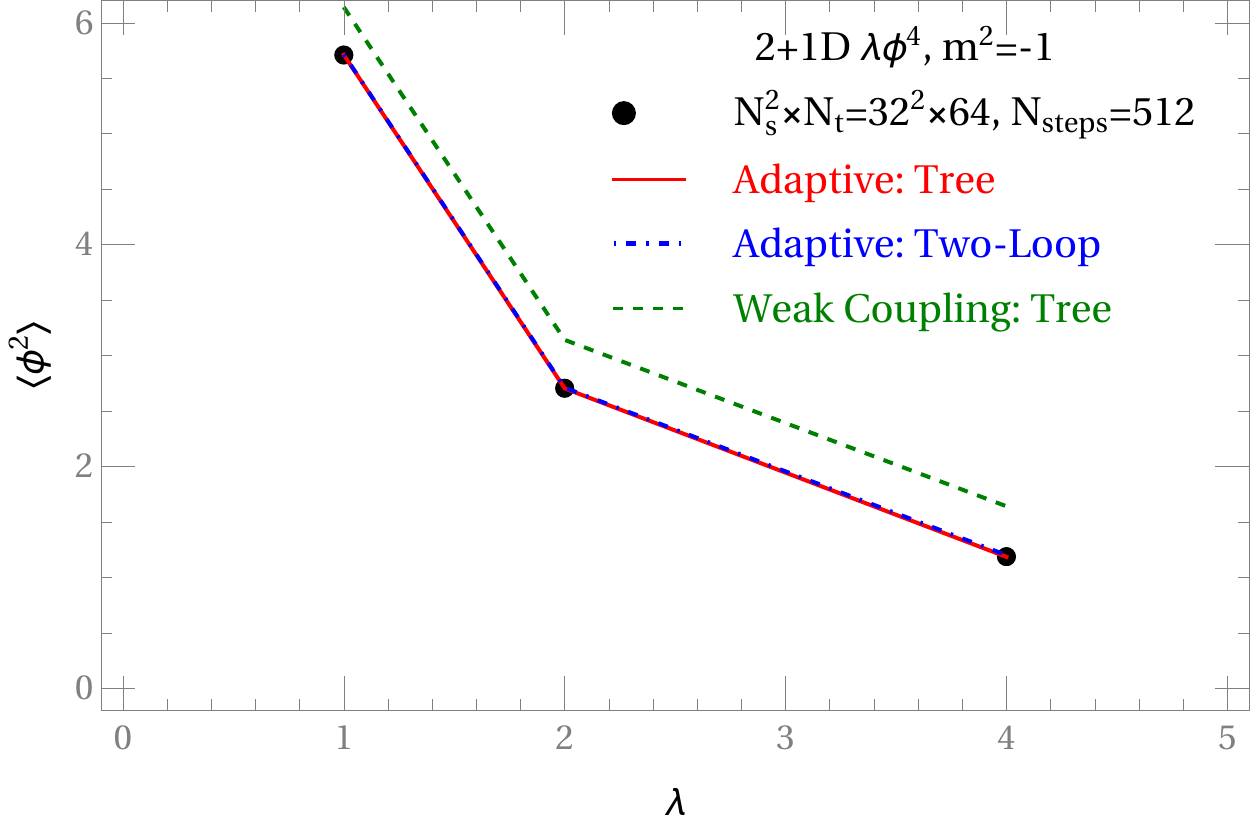}
    \includegraphics[width=0.49\textwidth]{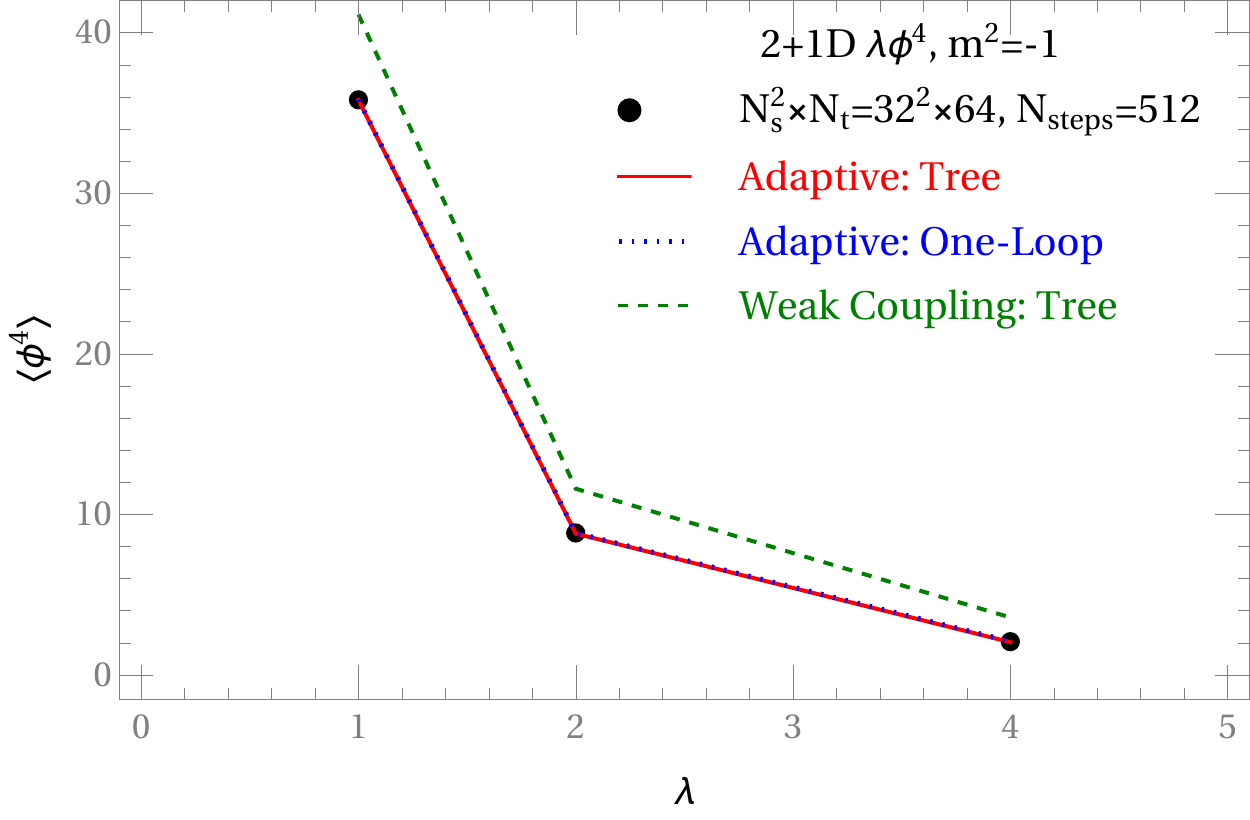}
\end{center}
\caption{We compare the perturbation result to the Hybrid Monte Carlo Simulation with the fixed bare mass ($m^2=-1$) and different values of bare coupling constant ($\lambda=1, 2, 4$). 
The result of the adaptive perturbation is also compared to the weak coupling expansion at the tree level. 
The number of measurements is $2^{6}$ sweeps with thermalization $2^8$ sweeps and measure intervals $2^6$ sweeps. 
The error bars are less than $1\%$. 
We choose the spatial size $N_s=32$ and temporal size $N_t=64$ in the lattice simulation. 
The number of molecular dynamics steps is $N_{\mathrm{steps}}=512$. 
The perturbation order is up to two-loop for $\langle\phi\rangle$ and $\langle\phi^2\rangle$. 
For $\langle\phi^4\rangle$, the order is up to one-loop. 
When $\phi_0\neq0$, spontaneous symmetry breaking \cite{Goldstone:1961eq} happens. 
We simulate two vacuum states on the lattice and compute $\langle|\phi|\rangle$ rather than $\langle\phi\rangle$. 
}
\label{Lattice_Ana_4}
\end{figure}

\noindent 
We replace the propagator with the lattice propagator as in the following
\bea
&&
\int\frac{d^3k_E}{(2\pi)^3}\ \frac{1}{k_E^2+\gamma^2(k_E)}
\nn\\
&\rightarrow& 
\int\frac{d^3k_E}{(2\pi)^3}\ \frac{1}{\gamma^2(k_E)+\sum_{\mu=1}^3(2-2\cos(k_{E, \mu}))}  
\eea
for comparing the perturbation result to the lattice simulation. 
We choose the unit lattice spacing on the lattice. 
We use the result of Ref. \cite{Ma:2022atx} to obtain $\langle\phi\rangle$ (two-loop), $\langle\phi^2\rangle$ (two-loop), and $\langle\phi^4\rangle$ (one-loop). 
The perturbation results match well with the lattice simulation as in Fig. \ref{Lattice_Ana_4}.
The advantage of the adaptive perturbation method can be found by comparing it to weak coupling expansion at the tree-level diagram in Fig. \ref{Lattice_Ana_4}. 
Since a finite lattice cannot have spontaneous symmetry breaking \cite{Goldstone:1961eq}, we compute $\langle|\phi|\rangle$ from the Hybrid Monte Carlo simulation when $\phi_0\neq 0$. 

\subsection{$|\phi|^3$ Theory}
\noindent 
Another asymptotic free scalar field theory is the (5+1)d $\lambda\phi^3$ theory. 
The Lagrangian is that replaces the $\phi^4$ theory's four-scalar interaction with $\lambda\phi^3/3!$. 
This theory does not have a stable vacuum. 
It is hard to have a reliable result from the Monte Carlo simulation. 
Hence we consider the $|\phi|^3$ theory. 
The Lagrangian has $\mathbb{Z}_2$ symmetry ($\phi\leftrightarrow-\phi$). 
Therefore, the $|\phi|^3$ theory has a stable vacuum. 
When considering a small enough $\lambda$ in $\phi^3$ theory, the fluctuation is also around the stable vacuum. 
Therefore, $|\phi|^3$ theory should provide an approximate description of $\phi^3$ theory at high energies. 
The similar high-energy description also implies that $|\phi|^3$ theory is asymptotic freedom. 
Hence we are interested in implementing the adaptive perturbation method to $|\phi|^3$ theory and comparing the result to the lattice simulation.  
\\

\begin{figure}
\begin{center}
\includegraphics[width=1.\textwidth]{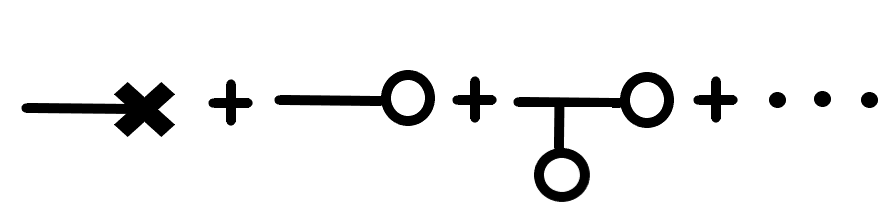}
\end{center}
\caption{We show the resummation of tadpole diagrams up to the two loops.}
\label{one-point}
\end{figure}

\begin{figure}
\begin{center}
    \includegraphics[width=0.49\textwidth]{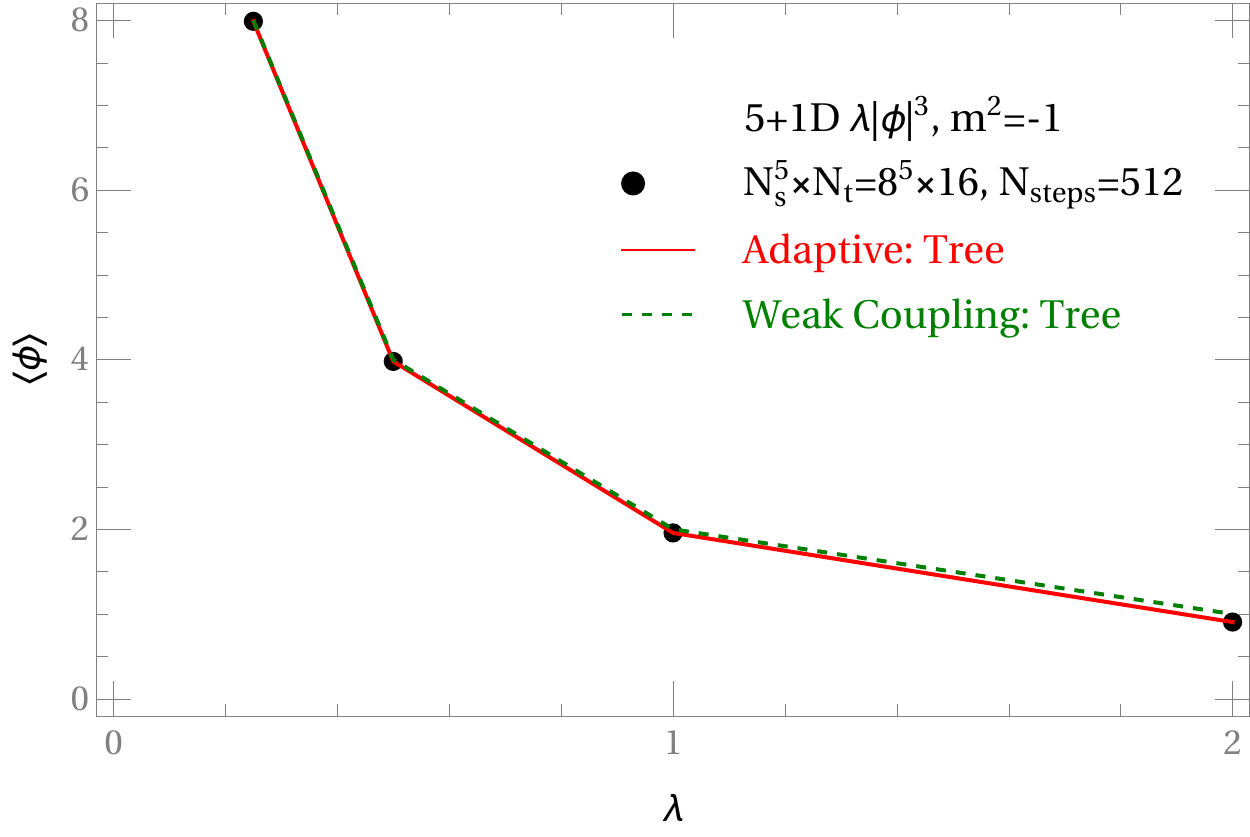}
    \includegraphics[width=0.49\textwidth]{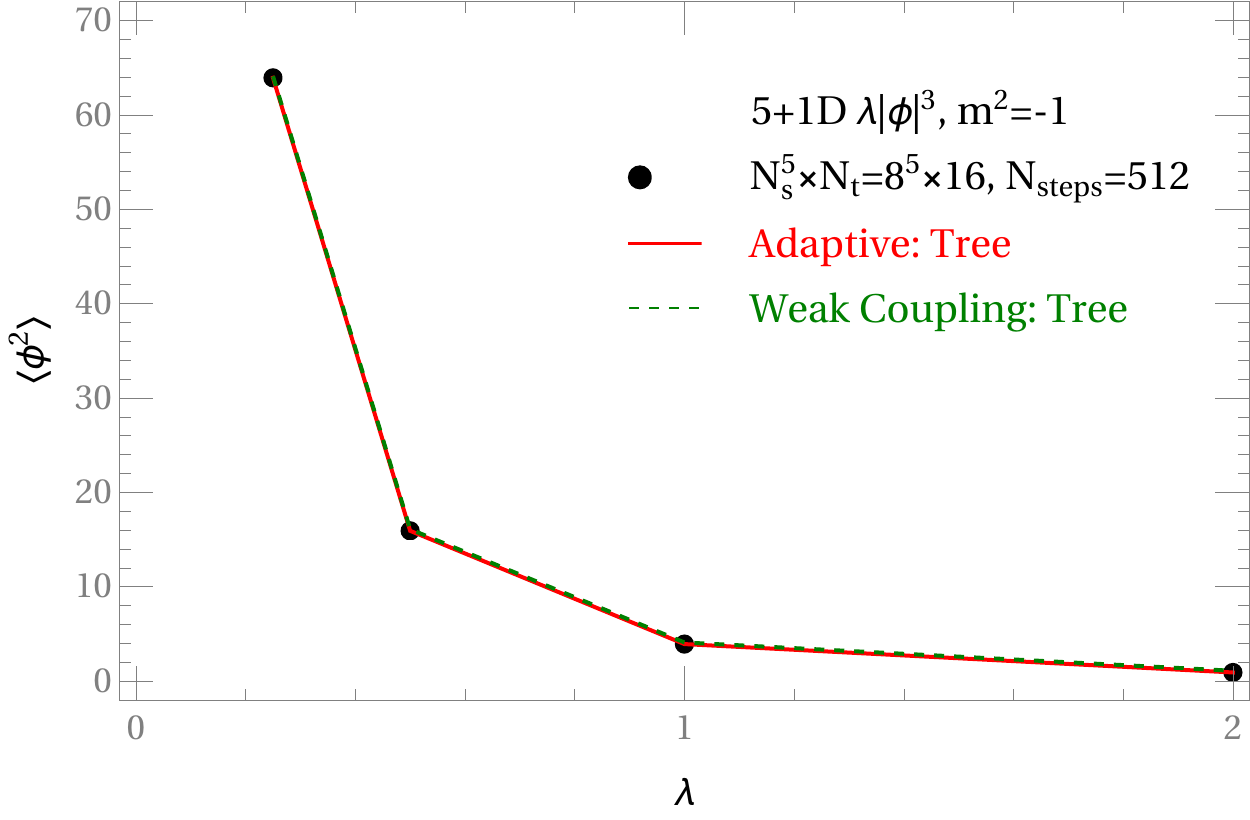}
\end{center}
\caption{We compare the perturbation result to the Hybrid Monte Carlo Simulation with the fixed bare mass ($m^2=-1$) and different values of bare coupling constant ($\lambda=0.25, 0.5, 1, 2$). 
The result of the adaptive perturbation is also compared to the weak coupling expansion at the tree level. 
The number of measurements is $2^{9}$ sweeps with thermalization $2^9$ sweeps and measure intervals $2^8$ sweeps. 
The error bars are less than $1\%$. 
We choose the spatial size $N_s=8$ and temporal size $N_t=16$ in the lattice simulation. 
The number of molecular dynamics steps is $N_{\mathrm{steps}}=512$. 
The perturbation order is up to tree diagrams for $\langle\phi\rangle$ and $\langle\phi^2\rangle$. 
}
\label{Lattice_Ana_3}
\end{figure}

\noindent 
We can consider $|\phi|^3$ theory as $\phi^3$ theory with non-negative field values ($\phi\ge 0$). 
The convenient derivation of the Feynman rule is from Wick's theorem. 
We cannot apply Wick's theorem due to the loss of negative values on $\phi$. 
The perturbation calculation is doable but troublesome. 
Therefore, we only show the result up to the leading order. 
We first calculate the vacuum energy density 
\bea
&&
\frac{\langle 0| H_3|0\rangle}{V}
\nn\\
&=&
\frac{m^2}{2}\phi_0^2
+\frac{\lambda}{3!}\phi_0^3
+\int\frac{d^5\vec{k}}{(2\pi)^5}\ \bigg(\frac{\gamma(\vec{k})}{4}
+\frac{\vec{k}\cdot\vec{k}+m^2}{4\gamma(\vec{k})}
\bigg)
\nn\\
&&
+\frac{\lambda}{4}\phi_0\int\frac{d^5\vec{k}}{(2\pi)^5}\ \frac{1}{\gamma(\vec{k})}, 
\eea
where $V$ is the spatial volume. 
The extreme value of the vacuum energy density for $\gamma(\vec{k})$ requires that 
\bea
\gamma^2(\vec{k})=\vec{k}\cdot\vec{k}+\gamma^2(0), \ 
\gamma^2(0)=m^2+\lambda\phi_0.
\eea 
We choose $\phi_0$ by requiring the extreme value of the vacuum energy density for $\phi_0$. 
The $\phi_0$ satisfies that  
\bea
m^2\phi_0+\frac{\lambda}{2}\phi_0^2
+\frac{\lambda}{4}\int\frac{d^5\vec{k}}{(2\pi)^5}\ \frac{1}{\gamma(\vec{k})}=0.
\eea 
The minimum value of the vacuum energy density implies $\gamma^2(0)\ge 0$. 
Hence we obtain the unique solution of $\phi_0$, 
\bea
\phi_0=-\frac{m^2}{\lambda}+\sqrt{\frac{m^4}{\lambda^2}-\int\frac{d^6 k_E}{(2\pi)^6}\ \frac{1}{k_E^2+\gamma^2(0)}}. 
\eea 
Because this theory does not have a four-scalar interaction, the contribution of self-energy diagrams does not go into $\gamma^2(0)$. 
Expanding $\phi_0$ in terms of $\lambda$ shows the resummation of tadpole diagrams as in Fig. \ref{one-point}. 
The resummation result of $\langle\phi\rangle$ and $\langle\phi^2\rangle$ for $\lambda=0.25, 0.5, 1, 2$ matches the simulation result in Fig. \ref{Lattice_Ana_3}. 
The advantage of resummation can be found by comparing it to weak coupling expansion from $\langle\phi\rangle$ by increasing $\lambda$ in Fig. \ref{Lattice_Ana_3}. 
The adaptive perturbation method works well with the asymptotic free scalar field theories. 
Indeed, the three-vertex diagrams have a simple structure. 
Therefore, we can improve the resummation techniques to introduce the self-energy and vertex diagrams to the renormalized mass and coupling constant.  
    
\section{Improvement in $\lambda\phi^3$} 
\label{sec:3}
\begin{figure}
\begin{center}
\includegraphics[width=1.\textwidth]{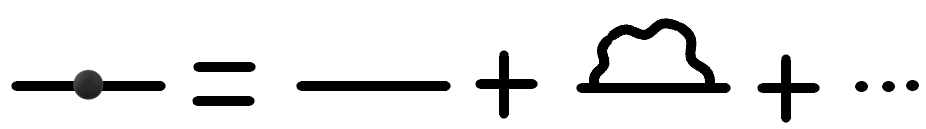}
\includegraphics[width=1.\textwidth]{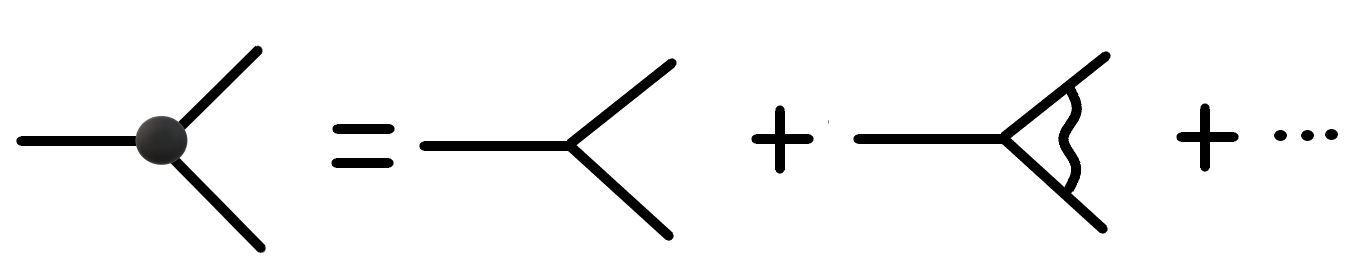}
\end{center}
\caption{We show the resummation of the self-energy and vertex diagrams up to the one-loop. The black dot indicates the resummation. }
\label{two-point and three-point}
\end{figure}

\noindent 
We propose the resummation of three-vertex diagrams from the $\phi^3$ theory because we are familiar with the Feynman rule. 
The one-point diagram sums all tadpole diagrams as in Fig. \ref{one-point}. 
We consider the resummation of self-energy diagrams to introduce the modified propagator as in Fig. \ref{two-point and three-point}. 
The resummation of the vertex diagrams gives the renormalized coupling constant $\lambda_R$ as in Fig. \ref{two-point and three-point}. 
Hence the constant condensation $\phi_0$, renormalized mass $\gamma(k_E)$, and renormalized coupling constant $\lambda_R(k_{1, E}, k_{2, E})$ become that: 
\bea
&&
\phi_0
\nn\\
&=&\frac{\gamma^2(0)}{\lambda_R(0, 0)}
\nn\\
&&
+\frac{\gamma^2(0)}{\lambda_R(0, 0)}\sqrt{1-\frac{\lambda_R(0, 0)}{\gamma^4(0)}\int\frac{d^6 k_E}{(2\pi)^6}\ \frac{\lambda_R(k_E, k_E)}{k_E^2+\gamma^2(k_E)}}; 
\nn\\
&&
\gamma^2(k_E)
\nn\\
&=&
m^2+\lambda_R(k_E, k_E)\phi_0
\nn\\
&&
-\int\frac{d^6q_E}{(2\pi)^6}\ \frac{\lambda_R^2(k_E-q_E, q_E)}{2}
\frac{1}{q_E^2+\gamma^2(q_E)}
\nn\\
&&\times\frac{1}{(k_E-q_E)^2+\gamma^2(k_E-q_E)}; 
\eea
\bea 
\lambda_R(k_{1, E}, k_{2, E})=\lambda+F(k_{1, E}, k_{2, E}), 
\eea
where 
\bea
&&
F(k_{1, E}, k_{2, E})
\nn\\
&\equiv&
\int\frac{d^6q_{1, E}}{(2\pi)^6}\ \lambda_R(k_{1, E}, k_{1, E}-q_{1, E})
\nn\\
&&\times
\lambda_R(k_{1, E}-q_{1, E}, k_{2, E}-q_{1, E})
\lambda_R(k_{2, E}-q_{1, E}, k_{2, E})
\nn\\
&&\times
\frac{1}{q_{1, E}^2+\gamma^2(q_{1, E})}
\frac{1}{(k_{1, E}-q_{1, E})^2+\gamma^2(k_{1, E}-q_{1, E})}
\nn\\
&&\times
\frac{1}{(k_{2, E}-q_{1, E})^2+\gamma^2(k_{2, E}-q_{1, E})}. 
\eea 
The $\lambda_R$ has the permutation symmetry for the momenta, 
\bea
\lambda_R(k_{1, E}, k_{2, E})=\lambda_R(k_{2, E}, k_{1, E}).
\eea 
The resummation of self-energy and vertex diagrams introduces the momentum dependence to the renormalized mass. 
The strongly coupled effect should change the pole of the propagator. 
The flow of $m^2$ also helps analyze the resummation contribution.  
Hence we are interested in investigating the theoretical properties from the RG flow \cite{Gell-Mann:1954yli} and the resummation formula. 

\section{RG Flow}
\label{sec:4}
\noindent  
We introduce the momentum cut-off $\Lambda$ to regularize the integration. 
The pole of a propagator determines the physical mass  $m_p$. 
We choose the physical coupling constant $\lambda_p$ from the renormalized coupling constant $\lambda_R$ when each square of momenta is $m_p^2$. 
We then calculate the running of bare parameters in (2+1)d $\phi^4$ and (5+1)d $\phi^3$ theories. 

\subsection{$\phi^4$ Theory}
\begin{figure}
\begin{center}
\includegraphics[width=0.49\textwidth]{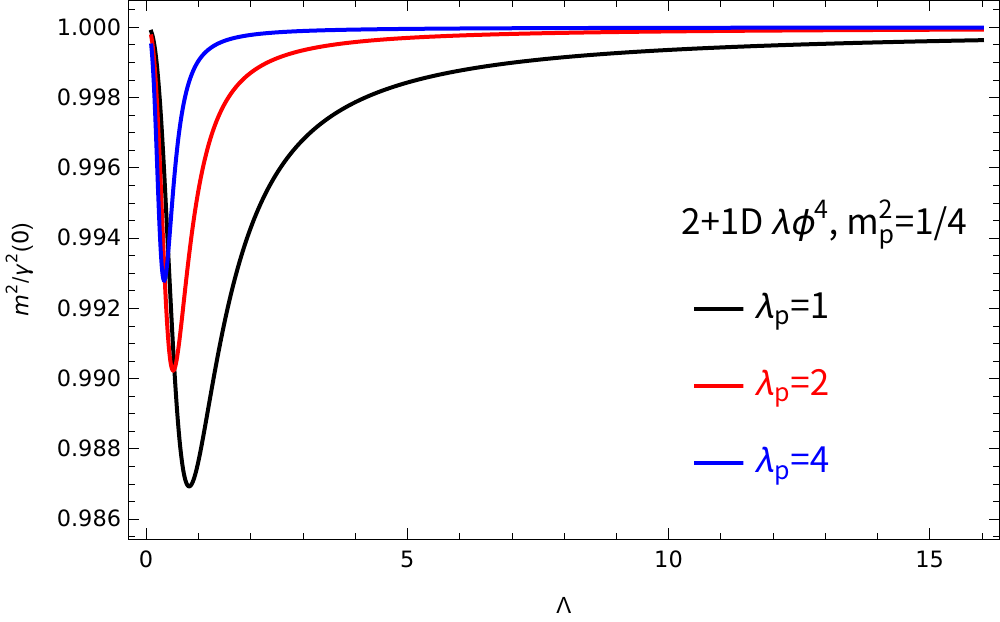}
\includegraphics[width=0.49\textwidth]{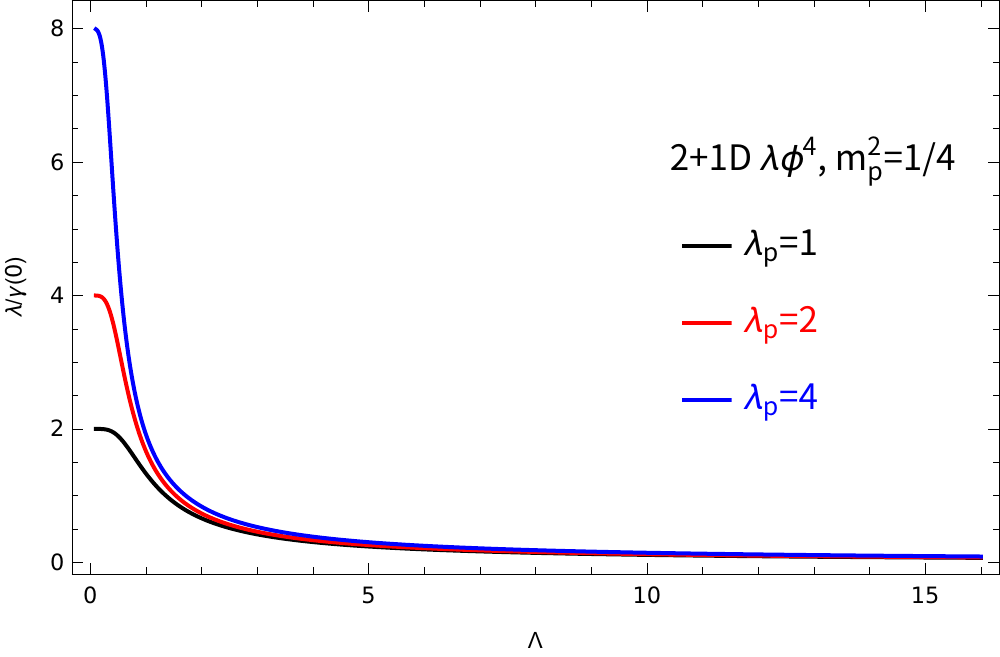}
\end{center}
\caption{We choose $\phi_0=0, m_p^2=1/4, \lambda_p=1, 2, 4$ in (2+1)d $\lambda\phi^4$ theory. 
The flow of $\lambda/\gamma(0)$ shows the asymptotic freedom. 
The $m^2/\gamma^2(0)$ is near one during the RG flow. 
}
\label{phi4}
\end{figure}

\noindent 
We choose $\phi_0=0, m_p^2=1/4, \lambda_p=1, 2, 4$ and calculate the dimensionless variables $m^2/\gamma^2(0)$ and $\lambda/\gamma(0)$ for $\Lambda$ in Fig. \ref{phi4}.
When $\Lambda$ becomes larger, the $\lambda$ flows to zero. 
Therefore, the (2+1)d $\lambda\phi^4$ theory is asymptotic free.   
The variable $m^2/\gamma^2(0)$ is almost one during the RG flow. 

\subsection{$\phi^3$ Theory}
\begin{figure}
\begin{center}
\includegraphics[width=0.49\textwidth]{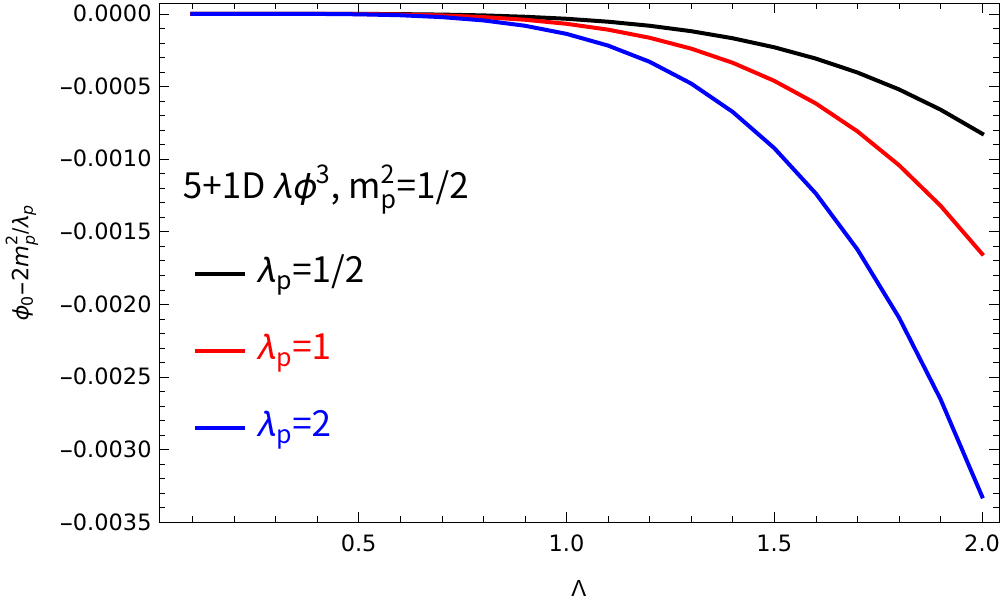}
\includegraphics[width=0.49\textwidth]{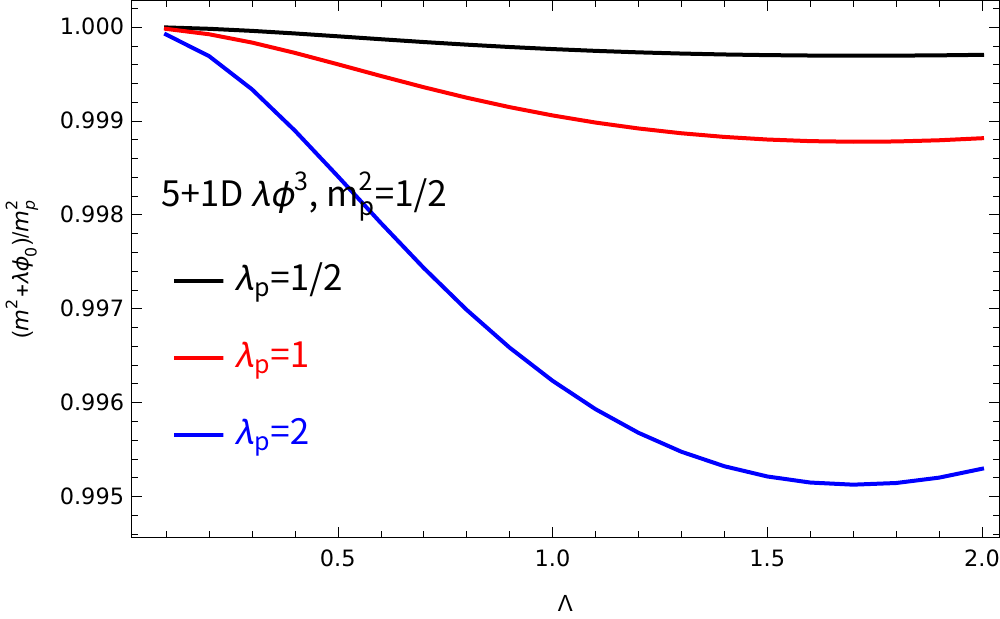}
\includegraphics[width=0.49\textwidth]{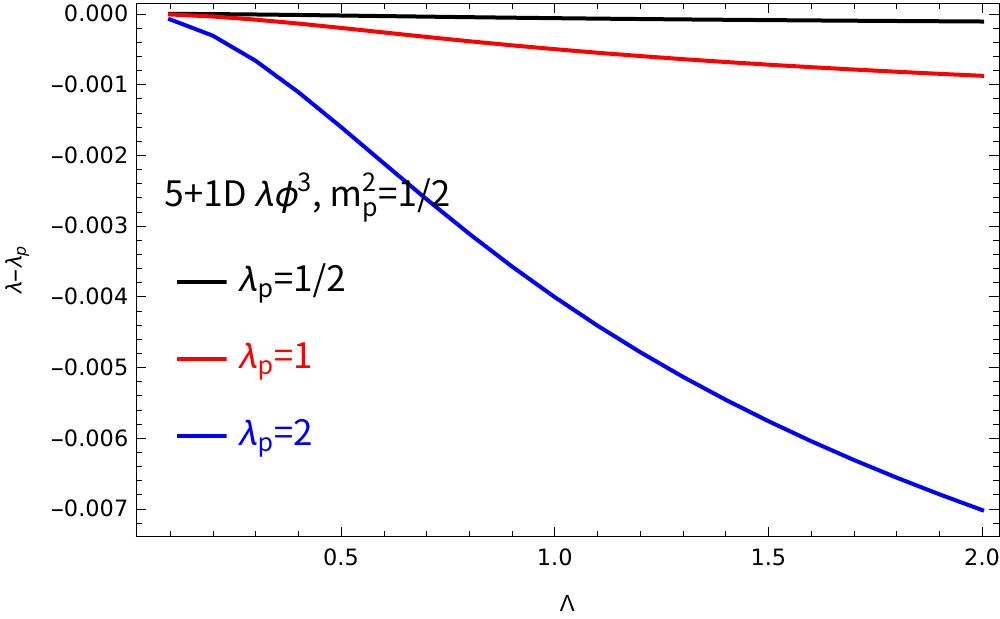}
\end{center}
\caption{We choose $m_p^2=1/2, \lambda_p=0.5, 1, 2$ in (5+1)d $\lambda\phi^3$ theory. 
The flow of $\lambda$ shows the asymptotic freedom. 
The flow of $(m^2+\lambda\phi_0)/m_p^2$ has a more deviation from 1 for a stronger coupling (large $\lambda_p$). 
}
\label{phi3}
\end{figure}

\noindent 
We choose $\phi_0\neq 0, m_p^2=1/2, \lambda_p=0.5, 1, 2$ and calculate the variables $\phi_0$, $(m^2+\lambda\phi_0)/m_p^2$, and $\lambda$ for $\Lambda$ in Fig \ref{phi3}. 
The running of $\lambda$ is similar to the (2+1)d $\lambda\phi^4$ theory. 
Therefore, the (5+1)d $\phi^3$ theory is also asymptotic free.   
When the energy scale approaches infinite, the bare coupling constant and condensation should flow to zero. 
At a high enough energy scale, the coupling constant is small enough to let the fluctuation be around the stable vacuum. 
Hence the RG flow supports that the $|\phi|^3$ theory provides a high-energy description of the $\phi^3$ case. 
We also observe that the flow of $(m^2+\lambda\phi_0)/m_p^2$ has more fluctuations for a larger $\lambda_p$ (the situation is different from $\phi^4$ because $m^2/\gamma^2(0)$ does not decrease for $\Lambda$). 
This observation indicates that the strong interaction should significantly change the pole of the propagator or the physical mass. 
  
\section{Outlook}
\label{sec:5}
\noindent 
Our study of the adaptive perturbation method or resummation techniques \cite{Weinstein:2005kw,Weinstein:2005kx,Ma:2022atx} in various scalar field theories, including the (2+1)d $\lambda\phi^4$ theory and the (5+1)d $\lambda\phi^3$ or $\lambda|\phi|^3$ theories, has provided valuable insights into the behavior of these theories. 
We found that the adaptive perturbation method provided an accurate comparison to lattice simulations in the correlation function, and we improved the resummation technique in the $\phi^3$ theory to include a renormalized mass that depends on momentum. 
Our analysis of the RG flow revealed that the resummation contribution should be unavoidable in the (5+1)d $\phi^3$ theory and that the $|\phi|^3$ theory should give a high-energy description to the $\phi^3$ case in (5+1)d. 
We also suggest that our resummation techniques have broad applicability, particularly in exploring the tri-critical point in the (3+1)d Gross-Neveu-Yukawa model, which could have implications for a similar interest in QCD. 
Overall, our study has provided valuable insights into the behavior of these scalar field theories, especially in terms of the effectiveness of the adaptive perturbation method and the potential to improve our theoretical understanding. 
\\

\noindent 
Quantum triviality in 4D scalar fields \cite{Luscher:1987ay} suggests that a 4D scalar field theory, like a $\phi^4$ theory, becomes trivial at high energy scales. 
In this context, "trivial" means that the interactions effectively disappear, and the theory behaves like a free scalar field theory. 
This is a significant problem when studying scalar fields in 4D, as it implies that interactions cannot survive in the ultraviolet limit. 
\\

\noindent
Ref. \cite{Romatschke:2023sce} to a negative bare coupling constant offers an alternative way to avoid the quantum triviality issue.
The bare coupling constant refers to the value of the interaction strength before renormalization, a fundamental process in QFT that deals with infinities that arise during calculations.
Although the bare coupling is negative, the renormalized or physical coupling constant remains positive, preserving physical realism. The 4D scalar field theory directly impacts the Higgs boson in the Standard Model, which is modeled as a scalar field.
Studying the Higgs with this approach might help resolve issues related to triviality and asymptotic freedom.
Our suggestion of applying the computation method to the Higgs sector of the Standard Model is intriguing, as it could lead to a more accurate understanding of how the Higgs behaves at very high energies.
This might help clarify the triviality problem and offer insights into new physics beyond the Standard Model, particularly in the context of QFT and Renormalization.
  
\section*{Acknowledgments}
\noindent
We thank Shu-Yu Ho for his helpful discussion. 
CTM thanks Nan-Peng Ma for his encouragement.
CTM acknowledges the YST Program of the APCTP;   
China Postdoctoral Science Foundation, Postdoctoral General Funding: Second Class (Grant No. 2019M652926); 
Foreign Young Talents Program (Grant No. QN20200230017).  
HZ acknowledges the Guangdong Major Project of Basic and Applied Basic Research (Grant No. 2020B0301030008), 
the Science and Technology Program of Guangzhou (Grant No. 2019050001), 
and the National Natural Science Foundation of China (Grant No. 12047523 and 12105107). 




\baselineskip 22pt
\begin{thebibliography}{99}
\bibitem{Gell-Mann:1954yli}
M.~Gell-Mann and F.~E.~Low,
``Quantum electrodynamics at small distances,''
Phys. Rev. \textbf{95}, 1300-1312 (1954)
doi:10.1103/PhysRev.95.1300

\bibitem{Gross:1973id}
D.~J.~Gross and F.~Wilczek,
``Ultraviolet Behavior of Nonabelian Gauge Theories,''
Phys. Rev. Lett. \textbf{30}, 1343-1346 (1973)
doi:10.1103/PhysRevLett.30.1343

\bibitem{Politzer:1973fx}
H.~D.~Politzer,
``Reliable Perturbative Results for Strong Interactions?,''
Phys. Rev. Lett. \textbf{30}, 1346-1349 (1973)
doi:10.1103/PhysRevLett.30.1346

\bibitem{Luscher:1987ay}
M.~Luscher and P.~Weisz,
``Scaling Laws and Triviality Bounds in the Lattice phi**4 Theory. 1. One Component Model in the Symmetric Phase,''
Nucl. Phys. B \textbf{290}, 25-60 (1987)
doi:10.1016/0550-3213(87)90177-5

\bibitem{Aizenman:1981du}
M.~Aizenman,
``Proof of the Triviality of phi**4 in D-Dimensions Field Theory and Some Mean Field Features of Ising Models for D\ensuremath{>}4,''
Phys. Rev. Lett. \textbf{47}, 1-4 (1981)
doi:10.1103/PhysRevLett.47.1

\bibitem{Goldstone:1961eq}
J.~Goldstone,
``Field Theories with Superconductor Solutions,''
Nuovo Cim. \textbf{19} (1961), 154-164
doi:10.1007/BF02812722

\bibitem{Feynman:1963uxa}
R.~P.~Feynman, R.~B.~Leighton and M.~Sands,
``The Feynman Lectures on Physics,''

\bibitem{Ma:2018efs}
C.~T.~Ma,
``Parity Anomaly and Duality Web,''
Fortsch. Phys. \textbf{66}, no.8-9, 1800045 (2018)
doi:10.1002/prop.201800045
[arXiv:1802.08959 [hep-th]].

\bibitem{Hioe:1978jj}
F.~T.~Hioe, D.~Macmillen and E.~W.~Montroll,
``Quantum Theory of Anharmonic Oscillators: Energy Levels of a Single and a Pair of Coupled Oscillators with Quartic Coupling,''
Phys. Rept. \textbf{43}, 305-335 (1978)
doi:10.1016/0370-1573(78)90097-2

\bibitem{Weinstein:2005kw}
M.~Weinstein,
``Adaptive perturbation theory. I. Quantum mechanics,''
[arXiv:hep-th/0510159 [hep-th]].

\bibitem{Weinstein:2005kx}
M.~Weinstein,
``Adaptive perturbation theory: Quantum mechanics and field theory,''
Nucl. Phys. B Proc. Suppl. \textbf{161}, 238-247 (2006)
doi:10.1016/j.nuclphysbps.2006.08.059
[arXiv:hep-th/0510160 [hep-th]].

\bibitem{Bargmann:1962zz}
V.~Bargmann,
``On the Representations of the Rotation Group,''
Rev. Mod. Phys. \textbf{34} (1962), 829-845
doi:10.1103/RevModPhys.34.829

\bibitem{Ma:2019pxd}
C.~T.~Ma,
``Adaptive Perturbation Method in Quantum Mechanics,''
IOP SciNotes \textbf{2}, no.3, 035202 (2021)
doi:10.1088/2633-1357/ac12ba
[arXiv:1911.08211 [quant-ph]].

\bibitem{Ma:2020ipi}
C.~T.~Ma,
``Second-Order Perturbation in Adaptive Perturbation Method,''
JHAP \textbf{2}, no.2, 37 (2022)
[arXiv:2004.00842 [hep-th]].

\bibitem{Ma:2020syr}
C.~T.~Ma,
``Accurate Study from Adaptive Perturbation Method,''
Int. J. Mod. Phys. A \textbf{36}, no.04, 2150029 (2021)
doi:10.1142/S0217751X21500299
[arXiv:2007.09080 [hep-th]].

\bibitem{Ma:2022atx}
C.~T.~Ma, Y.~Pan and H.~Zhang,
``Explore the Origin of Spontaneous Symmetry Breaking from Adaptive Perturbation Method,''
JHAP \textbf{4}, no.1, 51-64 (2024)
doi:10.22128/jhap.2024.754.1066
[arXiv:2205.00414 [hep-th]].

\bibitem{Andersen:2002ey}
J.~O.~Andersen, E.~Braaten, E.~Petitgirard and M.~Strickland,
``HTL perturbation theory to two loops,''
Phys. Rev. D \textbf{66}, 085016 (2002)
doi:10.1103/PhysRevD.66.085016
[arXiv:hep-ph/0205085 [hep-ph]].

\bibitem{Schwinger:1951ex}
J.~S.~Schwinger,
``On the Green's functions of quantized fields. 1.,''
Proc. Nat. Acad. Sci. \textbf{37}, 452-455 (1951)
doi:10.1073/pnas.37.7.452

\bibitem{Symanzik:1973hx}
K.~Symanzik,
``A field theory with computable large-momenta behavior,''
Lett. Nuovo Cim. \textbf{6S2}, 77-80 (1973)
doi:10.1007/BF02788323

\bibitem{Kleefeld:2005hf}
F.~Kleefeld,
``Kurt Symanzik: A Stable fixed point beyond triviality,''
J. Phys. A \textbf{39}, L9-L16 (2006)
doi:10.1088/0305-4470/39/1/L02
[arXiv:hep-th/0506142 [hep-th]].

\bibitem{Romatschke:2023sce}
P.~Romatschke,
``What if \ensuremath{\phi}4 theory in 4 dimensions is non-trivial in the continuum?,''
Phys. Lett. B \textbf{847}, 138270 (2023)
doi:10.1016/j.physletb.2023.138270
[arXiv:2305.05678 [hep-th]].


                                        
\end{thebibliography}
\end{document}